\documentclass[twocolumn,showpacs,preprintnumbers,amsmath,amssymb,prl]{revtex4}
\usepackage{graphicx}%  Include   figure  files  \usepackage{dcolumn}%
\begin{document}

% \preprint{APS/123-QED}

\title{Light-matter interaction in Mie nanolasers}

\author{C. Conti$^{1,2}$, A.  Fratalocchi$^{1,2}$ and G. Ruocco$^{2,3}$}
\affiliation{  $^1$ Museo  Storico  per  la Fisica  e  Centro studi  e
Ricerche ``Enrico Fermi'', Via Panisperna 89/A, I-00184, Roma, Italy\\
$^2$ Research center CRS-SOFT, c/o University of Rome ``La Sapienza'',
I-00185, Rome, Italy \\
$^3$ Dipartimento di Fisica, University of Rome ``La Sapienza'', I-00185, Rome, Italy 
}

\date{\today}

\begin{abstract}
By deriving a three  dimensional vector set of Maxwell-Bloch equations,
we  report  on an  \emph{ab-initio}  investigation  of  a  spherical  Mie
nanolaser.  Parallel  numerical simulations  predict  a rich  physical
scenario, ranging from  a nontrivial vectorial energy-matter interplay
in the initial  stage of evolution to mode competition and
dynamical   frequency   pulling    effects. 
Application  of these effects  could favor the  realization of
nonlinearly-controlled  largely-tunable nanolaser devices.
\end{abstract}

\pacs{32.80.-t, 42.65.Sf, 42.55.-f}

%42.55.-f Lasers
%42.65.Sf Dynamics of nonlinear optical systems; optical instabilities,
%	 optical chaos and complexity, and optical spatio-temporal dynamics
%32.80.-t Photon interactions with atoms

\maketitle

\begin{figure}[h!]
\includegraphics[width=8.5 cm]{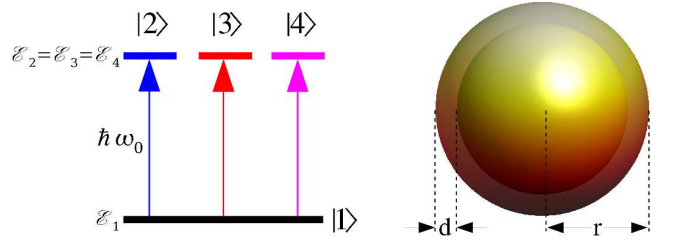}
\caption{\label{ed0}  (Color  online).  (left)  Energy  diagram  of  a
four-level  atomic  system  with  a triply-degenerate  excited  state;
(right) sketch of a single Mie nanoresonator with a $d-$thick layer of
amplifying material.  }
\end{figure}

\begin{figure}
\includegraphics[width=8.5 cm]{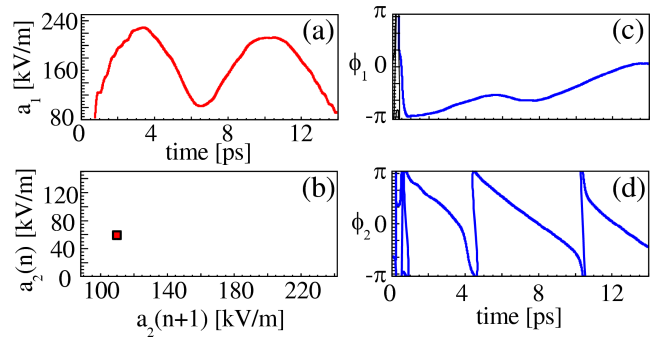}
\caption{\label{tw1}   (Color    online).   Frequency   demodulation
analysis: time evolution of modal amplitude $a_1(t)$ (a) and phases $\phi_{1,2}(t)$ (c-d); (b) 1D map obtained with a Poincar\'e surface of section of $(a_1,a_2)$ at $a_1=180$ kV/m.
  }
\end{figure}

%\paragraph{Introduction. ---}
Since  the pioneering investigations of  Gustav Mie  in the
early days of the  last century \cite{mie_th_mie}, the theoretical and
experimental study  of spherical  resonators has deserved  much attention
from  the scientific  community.  This interest  stems from the
analysis of  fundamental  processes such as  scattering,
energy  propagation  through   disordered  media  and  cavity  quantum
electrodynamics, and from the large number  of applications in
photonics,    chemistry,    meteorology,    astronomy   and    sensing
\cite{LSBSP,tm_mie_xua,SCQE,cqe_exp_yukawa,OEAWSP,TSOLAOER,mie_rs_wyatta,mie_ls_ashkia}. Mie
resonators  display  ultrahigh
quality  factors $Q$  $(\approx 10^9)$  \cite{mie_qf_vahala}, and are able to
confine  and store  electromagnetic energy  for long  times
within  small dielectric  volumes;
Mie whispering galleries modes (WGM) can sustain
enhanced stimulated emission processes
so   that    'thresholdless'   lasing   can   be   observed
\cite{mie_tl_skusha,mie_tl_vietz,mie_tl_spilla,mie_tl_sando}. 
Despite the widespread scientific production, the theoretical
understanding of  Mie lasers  is still an  open field of  research. 
For what concerns laser emission, existing theoretical approaches 
either rely on  a semi-classical treatment
based    on   light    interaction   with    a   two    level   system
\cite{lt_tl_jaynea,QN,ol_th_lamba} or on rate equations \cite{LASERS},
with  numerical  analysis limited  to  one and  two-dimensional
media  \cite{tl_fd_slavca}. 
The  former approach,  as was recently  observed \cite{fdmb_2d_slavc},  
is  not   rigorous  as  the  number  of  physical
dimensions  grows above  one   while  the latter,
accounting  only for  the contribution  of atomic  populations, misses
important informations on quantum  coherence
\cite{re_mb_bidega},
and  cannot be effectively pursued  to perform any
realistic  \emph{ab initio}  computation at ultra-fast time-scales ($<$ps). Furthermore, 
although  low dimensional models permit a simplified analysis with respect to the fully vectorial set
of Maxwell  equations they left  the general picture unknown.  
{\it Which is the outcome of a strongly nonlinear ultra-fast  and multi-dimensional
interaction on scales comparable with the wavelength?} 
Beside being central for the development of laser theory, this question has fundamental importance in quantum mechanics \cite{OCAQO}, 
soliton \cite{sit_op_mccala} and chaos theory \cite{OTT}.
However,  to the  best of our  knowledge, neither  \emph{ab
initio}  investigations  were  reported  nor  quantum-mechanical
models    of   energy-matter    interaction for three-dimensional  nanostructured lasing systems were developed. 
Owing to the large interest in the study of nano-scaled particle aggregates \cite{nl_th_kuirua,ld_th_contia,ph_pr_rojas}, 
the theoretical investigation of such  dynamics is of considerable practical importance.\\  
In  this Letter  we develop  an  \emph{ab  initio}  rigorous theoretical  model  of  light
interaction in the presence  of amplifying (or dissipative) materials,
deriving  a  three  dimensional   vector  set  of  Maxwell-Bloch  (MB)
equations  within the  real  representation in  terms  of the  $SU(n)$
algebra \cite{mb_th_hioea}. We  discretized the resulting equations on
a Yee  grid and numerically  solved them within  the Finite-Difference
Time-Domain  (FDTD) method  \cite{CETFDTDM}. The  MB-FDTD  approach is
then   applied  to   investigate   the  ultrafast   dynamics  of   Mie
nanoresonators.  Specifically, we perform  a  series of  numerical  experiments
by investigating the process of  laser emission from a single nanosphere,
covered  by  a  layer   of  active  material,  for  different  pumping
rates. This theoretical analysis yields the following scenario:\\ i) In
the  transient  before  the  steady  state  regime  strong
interactions  between all components of the electromagnetic field and  the atom  take
place. Such a \emph{vectorial} dynamics
is the key ingredient for the reaching of a stable steady state.\\ ii)
When  the  lasing  process  settles  up,  modes  possessing
frequencies  close  to  Mie  resonances get  excited  and  competition
dynamics begins.  
%Such effects are revealed on fields
%time   evolution  by   means  of   both  spectrograms   and  frequency
%demodulation.
\\  
iii) As  the  pumping  rate gets  higher,
a strongly multi-mode  regime occurs. In this  situation modes
with higher  quality factors tend to  grow faster than  the others and
dynamical frequency pulling effects are  observed. In particular, as the pumping
rate increases the principal  frequency of laser emission moves toward
smaller wavelengths encompassing large shifts; a process radically different from the standard nonlinear
frequency shift observed in lasers. The mechanism investigated here can be at the basis of the development of nonlinearly-controlled largely-tunable nanolaser devices.
\\
%\paragraph{Theory. ---}
We begin by writing Maxwell equations in an isotropic medium:
\begin{align}
\label{maxw0}
&\frac{\partial\mathbf{H}}{\partial     t}=-\frac{1}{\mu_0}\nabla\times
\mathbf{E},
%\nonumber\\
&\frac{\partial\mathbf{E}}{\partial
t}=\frac{1}{\epsilon_0}\bigg[\nabla\times
\mathbf{H}-\frac{\partial\mathbf{P}}{\partial t}\bigg],
\end{align}
being  $\mathbf{P}$  the  material  polarization. The  latter  can  be
decomposed     as    the    sum     of    a     linear    contribution
$\mathbf{P}_\mathrm{L}=\epsilon_0(\epsilon_r-1)\mathbf{E}$   plus  a   nonlinear  term
$\mathbf{P}_{\mathrm{NL}}=-eN_a\langle \hat{\mathbf{Q}}\rangle$ modeling     the     quantum-mechanical
interaction between the electromagnetic field and the atoms  
%\begin{align}
%\label{nt0}
%\mathbf{P}_\mathrm{NL}=-eN_a\langle \hat{\mathbf{Q}}\rangle,
%\end{align}
being $e$ the electric charge, $N_a$ the  density of polarizable
atoms  and $\langle  \hat{\mathbf{Q}}\rangle$ the  expectation  of the
displacement operator  $\hat{\mathbf{Q}}$ with respect  to the quantum
state   $\lvert\psi(\mathbf{r},t)\rangle$.
To   derive   a   quantum
mechanical  model of  the material,  we consider  a  four-level
system, with three degenerate levels, in  which three electric mutually orthogonal dipole transitions can be  excited by
linearly  polarized electromagnetic  waves  with energy  equal to  the
atomic  resonance  $\hbar\omega_0$  (Fig. \ref{ed0}-left).  The  total
Hamiltonian  operator $\hat{\mathbf{H}}$  of  the system  can be  then
expressed as $\hat{\mathbf{H}}=\hat{\mathbf{H}}_0+\hat{\mathbf{H}}_I$,
%\begin{align}
%  \hat{\mathbf{H}}=\hat{\mathbf{H}}_0+\hat{\mathbf{H}}_I
%\end{align}
 with the  unperturbed Hamiltonian $\hat{\mathbf{H}}_0$ of components
$H_{ij}=\hbar\omega_0\delta_{i,j}(\delta_{i,2}+\delta_{i,3}+\delta_{i,4})$
($\delta_{i,j}$ is the Kronecker delta)
%\begin{align}
%  \hat{\mathbf{H}}_0=\hbar\omega_0\begin{bmatrix}
%0 &0 &0 &0\\
%0 &1 &0 &0\\
%0 &0 &1 &0\\
%0 &0 &0 &1
%\end{bmatrix},
%\end{align}
and        the       self-adjoint        interaction       Hamiltonian
$\hat{\mathbf{H}}_I=e\mathbf{E}\cdot\hat{\mathbf{Q}}$.  The displacement      operator   is    $\hat{\mathbf{Q}}=\hat{\mathbf{G}}+\hat{\mathbf{G}}^\dagger$ with
$G_{ij}=q_0\delta_{i,1}(\delta_{j,2}\mathbf{x}+\delta_{j,3}\mathbf{y}+\delta_{j,4}\mathbf{z})$,
%\begin{align}
%  \hat{\mathbf{Q}}=q_0\begin{bmatrix}
%0 &\mathbf{x} &\mathbf{y} &\mathbf{z}\\
%\mathbf{x} &0 &0 &0\\
%\mathbf{y} &0 &0 &0\\
%\mathbf{z} &0 &0 &0
%\end{bmatrix},
%\end{align}
being  $q_0$  the  typical   atomic  length  scale  and  $\mathbf{x}$,
$\mathbf{y}$, $\mathbf{z}$  unit vectors along $x$, $y$  and $z$ axes,
respectively.  The  dynamical  evolution   of  the  atomic  system  is
expressed by  the Liouville equation  of motion of the  density matrix
operator               $\hat{\rho}=\lvert\psi\rangle\langle\psi\lvert$:
$i\hbar\frac{\partial                              \hat{\rho}}{\partial
t}=[\hat{\mathbf{H}},\hat{\rho}]$. 
%\begin{align}
%\label{dm0}
%  i\hbar\frac{\partial \hat{\rho}}{\partial t}=[\hat{\mathbf{H}},\hat{\rho}],
%\end{align}
%with  $[\hat{A},\hat{B}]=\hat{A}\hat{B}-\hat{B}\hat{A}$  defining  the
%commutator between $\hat{A}$ and $\hat{B}$. 
By exploiting the symmetry
of  $\hat{\rho}$  under   the  $SU(4)$  group  \cite{mb_th_hioea},  we
introduce  the  coherence  vector  $\mathbf{S}$ of  components  $S_j$:
$S_j=\mathrm{Tr}[\hat{\rho}\hat{s}_j]$ $(j=1,2,...,15)$
% \begin{align}
%&   S_j=\mathrm{Tr}[\hat{\rho}\hat{s}_j], &j\in[1,15]
% \end{align}
being  Tr the trace  operator and  $\hat{s}_j$ the  $j-$th
generator  of   the  $SU(4)$  algebra   \cite{su4_bf_leea}.  The  time
evolution of the coherence $\mathbf{S}$ is then:
\begin{align}
\label{se0}
&                                    \frac{\partial\mathbf{S}}{\partial
t}=\hat{\Gamma}\mathbf{S}-\hat{\gamma}(\mathbf{S}-\mathbf{S}^{(0)}),
&\Gamma_{jh}=\frac{i}{2\hbar}\mathrm{Tr}\big(\hat{\mathbf{H}}[\hat{s}_j,\hat{s}_h]\big),
\end{align}
being $\hat{\gamma}$ a  phenomenologically added \cite{OCAQO} diagonal
matrix       of      nonuniform       relaxation       rates      with
$\gamma_{ii}=1/T_i$,  $\mathbf{S}^{(0)}$  a vector  containing
the  initial  populations of  the  system.  $\hat{\Gamma}$ is  skew and its nonzero components turn out to be:
% \begin{widetext}
% \begin{align}
% \label{se1}
% &\Gamma_{1,2}=\Gamma_{4,5}=\Gamma_{9,10}=\omega_0,
% &\Gamma_{1,7}=\Gamma_{2,6}=\Gamma_{3,5}=\frac{\Gamma_{8,5}}{\sqrt{3}}=\Gamma_{14,9}=\Gamma_{10,13}=\Omega_y,\nonumber\\
% &\frac{\Gamma_{3,2}}{2}=\Gamma_{7,4}=\Gamma_{5,6}=\Gamma_{12,9}=\Gamma_{10,11}=\Omega_x,
% &\Gamma_{1,12}=\Gamma_{2,11}=\Gamma_{3,10}=\Gamma_{4,14}=\Gamma_{5,13}=\frac{2\sqrt{2}\Gamma_{15,10}}{\sqrt{3}}=\Omega_z,
% \end{align}
% \end{widetext}
\begin{align}
\label{se1}
&\Gamma_{1,2}=\Gamma_{4,5}=\Gamma_{9,10}=\omega_0,\nonumber\\
&\Gamma_{1,7}=\Gamma_{2,6}=\Gamma_{3,5}=\frac{\Gamma_{8,5}}{\sqrt{3}}=\Gamma_{14,9}=\Gamma_{10,13}=\Omega_y,\nonumber\\
&\frac{\Gamma_{3,2}}{2}=\Gamma_{7,4}=\Gamma_{5,6}=\Gamma_{12,9}=\Gamma_{10,11}=\Omega_x,\nonumber\\
&\Gamma_{1,12}=\Gamma_{2,11}=\Gamma_{3,10}=\Gamma_{4,14}=\Gamma_{5,13}=\frac{2\sqrt{2}\Gamma_{15,10}}{\sqrt{3}}=\Omega_z,
\end{align}
with $\Omega_i=\frac{E_i\wp}{\hbar}$      ($i\in[x,y,z]$)     and
$\wp=eq_0$.  Finally,  the   expectation  value  of  the  displacement
operator $\hat{\mathbf{Q}}$ expressed in terms of $\mathbf{S}$ is:
\begin{align}
\label{exq0}
\langle\hat{\mathbf{Q}}\rangle=\mathrm{Tr}[\hat{\rho}\hat{\mathbf{Q}}]=q_0\big(\mathbf{x}
S_1+\mathbf{y} S_4+\mathbf{z} S_9\big).
\end{align}
Equations (\ref{maxw0})-(\ref{exq0})   
represent the three dimensional, full-wave vector set of Maxwell-Bloch
FDTD equations  modeling light-matter  interaction in the  presence of
resonant quantum transitions.\\
%\paragraph{Implementation ---}
Equations (\ref{se0})-(\ref{exq0}) have been discretized on
faces and sides of a cubic Yee grid while (\ref{maxw0}) on
its  center. Standard  leapfrogging  has been  used  as time  marching
algorithm for the FDTD,  second order Crank-Nicholson for the solution
of the MB  set. A Uniaxial Perfectly Matched Layer  (UPML) is employed to
absorb  outgoing waves.  Following \cite{tl_fd_slavca},  we  keep into
account fluctuations by adding to the electric field a Markovian noise
process  with   Gaussian  statistic,   the  latter  obtained   from  a
Box-M\"uller  random number  generator  algorithm. The  code has  been
parallelized within the Message-Passing-Interface (MPI) standard.\\ We
perform a  series of numerical experiments by  considering a spherical
resonator  (Fig.  \ref{ed0}-right)  of  r=$300$ nm  radius,
while taking $\epsilon_r=8.41$ as a realistic value for typical high-index materials
(e.g.  $TiO_2$ ),  covered by a  layer of active    material   of   thickness $d=100$   nm, with the same refractive index of the interior part and transition at
$\omega_0=3.88\cdot 10^{15}$  rad/s ($\lambda_0=483$ nm)
in proximity of Mie resonances, as detailed below.  By assuming
$q_0\approx 0.1$  nm, we  obtain a coupling  coefficient $\wp=4.8\cdot
10^{-28}$  $\mathrm{C\cdot m}$.  The  decaying constants  were set  to
$T_i=10$   fs   $(i=1,2,4,5,6,7,9,10,11,12,13,14)$   and  $T_j=1$   ns
$(j=3,8,15)$     according      to     the     existing     literature
\cite{tl_fd_slavca}. The pumping rate is settled 
by the density of polarizable atoms $N_a$ in the excited state,
which is assumed to be constant in time and fixed by an external continuous excitation 
involving additional energy levels, as for standard laser systems.
Furthermore, owing the small sphere radius, we assume a spatially uniform pumping
of the active layer.
The used discretization $\Delta x=\Delta y=\Delta
z=10$ nm and  $\Delta t=0.006$ fs guarantees both  high accuracy (with
$50$  points per wavelength at  $\lambda=500$ nm) as
well as temporal stability.
\\
\begin{figure}
\includegraphics[width=8.5 cm]{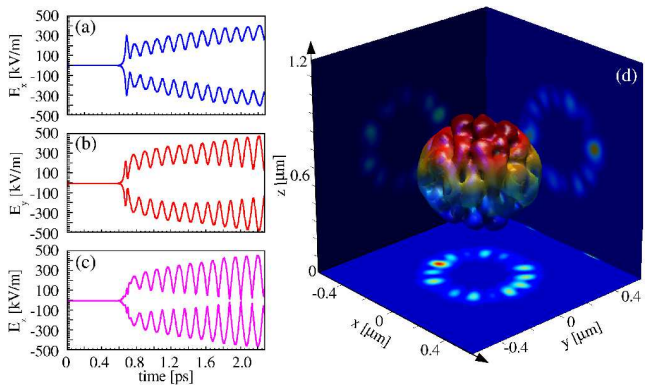}
\caption{\label{tra0}    (Color   online).    MB-FDTD    results   for
$N_a=10^{26}$ m$^{-3}$:  (a)-(c) time evolution of  the electric field
components (sampled in proximity of the sphere center, to avoid symmetry effects),  (d) energy density  $\mathcal{E}$ (isosurface  plot) with
$xy$ (down) $xz$  (left) and $yz$ (right) slices  in the sphere middle
plane.}
\end{figure}
%\paragraph{Transient regime.  ---}
We begin our analysis at a low pumping rate ($N_a=10^{26}$ m$^{-3}$). 
To investigate the transient regime, we collect at
each    time    step     all    the    electric    field    components
$E_x$, $E_y$, $E_z$,  sampled in  proximity of the  sphere center,   and    the   output   electromagnetic    energy   density
$\mathcal{E}=\mathbf{E}\cdot\mathbf{D}+\mathbf{H}\cdot\mathbf{B}$
(Fig. \ref{tra0}).
At $t\approx$ $600$ fs
the electromagnetic fields begin to coherently grow. Owing to the sphere-air high index contrast, the electromagnetic energy is well confined within the structure and the lasing mode experiences low losses (Fig. \ref{tra0}). This stage of evolution
is characterized  by a \emph{mutual} energy exchange  between the atom
and  each electromagnetic component  (Fig. \ref{tra0}a-c).  The latter
will  eventually interact  on the fs scale (fixed by $T_i=10$fs)  with each  other owing  to the  presence of
nonlinear  non   diagonal  coupling   terms  in  the   density  matrix
(\ref{se1}). From  a physical perspective,  energy exchange originates
from  the  presence  of  a  triply  degenerate  atomic  excited  state
(Fig.  \ref{ed0}, left panel), which  yields a  nontrivial  vectorial interplay
between the  electromagnetic field and the atom.  Such a contribution,
ignored by  previous theories,  is the
essential   ingredient for an isotropic laser three-dimensional dynamics.  
Without  a  four-level  atomic  system,  in  fact,  no
vectorial interaction is possible and no realistic steady state can be
theoretically predicted.\\
\begin{figure}
\includegraphics[width=8.5 cm]{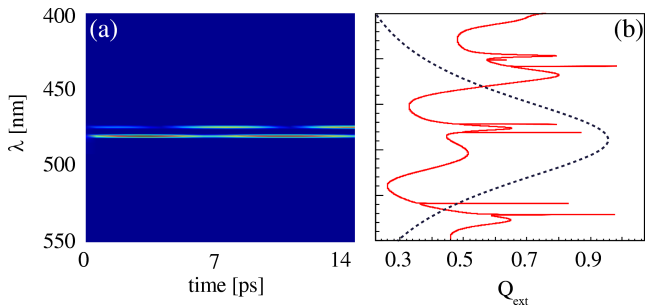}
\caption{\label{tw0}  (Color  online).  (a)  Time  series  spectrogram
computed  from  MB-FDTD   analysis  for  $N_a=10^{26}$  m$^{-3}$
(the mode-competition dynamics is denoted by the beating mode-amplitudes);
(b, solid line) Mie theory extinction factor $Q_{ext}$  for  a  single
nanosphere   of   diameter    $600$   nm   and   $\epsilon_r=8.41$; (b, dashed line)
active medium gain bandwidth.}
\end{figure}
%\paragraph{Two-mode competition dynamics. ---}
Once the transient stage  has been concluded ($t\approx 2$ ps), multimode lasing settles
up. Even if nano-sized, the sphere has high quality factors that do not prevent lasing to occur (Fig.   \ref{tw0}).
To study the time evolution of the steady state regime we make use
of   a   spectrogram   (Fig.   \ref{tw0}) and frequency demultiplexing (Fig. \ref{tw1}).   The   former   shows   a
\emph{mode-competition} dynamics between  two modes
in proximity of $\omega_0$. In Fig. \ref{tw0}b we plot the extinction factor $Q_{ext}$ of the nanosphere
as calculated from Mie theory  \cite{mie_th_mie}, which shows the available modes (Fig. \ref{tw0}b solid line) in the gain bandwidth (Fig. \ref{tw0}b dotted line). The  two  modes
 (as shown in  Figs. \ref{tw0}a-b) correspond  to the
nearest WGM modes with respect to the medium resonance $\omega_0$.
To characterize  the mode-competition dynamics,  we extract amplitudes
$a_1(t)$,  $a_2(t)$  and phases  $\phi_1(t)$,  $\phi_2(t)$ at  carrier
frequencies close to WGM modes  by means of frequency demodulation (Fig.  \ref{tw1}).   In  principle,   a   four  dimensional
phase-space $[a_1,a_2,\phi_1,\phi_2]$ supports a rich dynamics ranging
from periodic motion  to chaos. However, at this  low pumping rate, no
synchronization \cite{OTT}  occurs and each  phase independently linearly grows,
except in a small set of time instants where $a_i$ is
close to  zero (Fig. \ref{tw1}a-d). This is the ``free-run'' regime,
as originally predicted  by Lamb \cite{ol_th_lamba},
where the phase variables can be averaged out from the coupled-mode equations
and only the amplitudes remain as dynamical variables.
As a result,  we are left  with a
two-dimensional    phase    space     which,    according    to    the
Poincar\'e-Bendixson  theorem,  can   only  sustain  fixed  points  or
periodic motion,  as shown in Fig. \ref{tw1}a and in Fig. \ref{tw1}b, 
where we plot the map obtained from the Poincar\'e surface of section of $(a_1,a_2)$. The bi-modal steady state
is characterized by  a limit cycle where both  amplitudes $a_1$, $a_2$
are trapped together.\\
\begin{figure}
\includegraphics[width=8.5 cm]{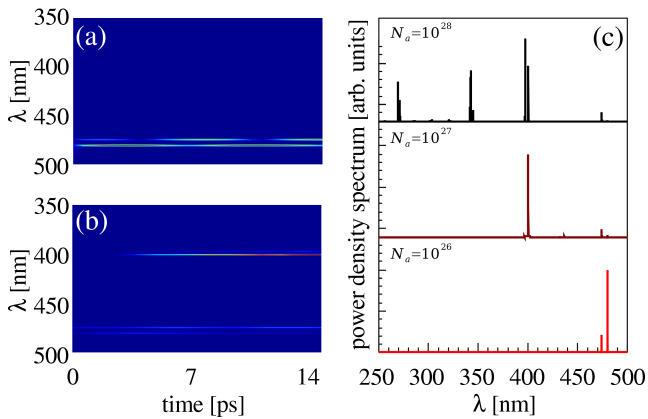}
\caption{\label{fp0}  (Color   online).  Spectrograms  for  increasing
pumping rates: $N_a=10^{26}$ m$^{-3}$ (a), $N_a=10^{27}$ m$^{-3}$; (c) Power density spectrum for various $N_a$.
 }
\end{figure}
%\paragraph{Dynamical frequency pulling effects in the multi-mode regime ---}
As the pumping rate $N_a$ gets higher, a large number of WGM
is above the threshold for laser oscillation. 
To  investigate the transition from two-mode
to multi-mode regime, we  calculate spectrograms obtained from MB-FDTD
simulations  for increasing  pumping rates  $N_a\in [10^{26},10^{28}]$
m$^{-3}$  (Fig.  \ref{fp0}).   The  dynamical  evolution  observed  is
characterized  by two  competing  effects. As  the frequency  $\omega$
increases with respect  to $\omega_0$: i) the gain  experienced by the
mode gets smaller; ii) the quality  factor $Q$ of the WGM mode becomes
higher. Therefore,  although low  wavelength  modes experience  small
gain,  they  tend to  grow  faster with  respect  to  modes at  higher
wavelengths.   The  result   of  this   competition  is   a
\emph{dynamical frequency  pulling} towards  smaller wavelength  as  the pumping
rate grows (Fig. \ref{fp0}c). The predicted frequency shift is very large ($\approx 100$ nm, see Fig. \ref{fp0}c) and quantitatively it depends on the 
\emph{morphology} of the nanosphere resonance spectrum 
(see Fig. \ref{tw0}b), which fix the involved wavelenghts and the corresponding Q-factors; 
in principle it can be 
controlled by acting on the shape of the resonator.
Dynamical frequency pulling is then expected to play a fundamental role in the nonlinear dynamics 
of laser systems, either based on a single sphere or on ordered and disordered resonator ensembles. 
The induced large shift can be exploited to develop largely-tunable nanosized laser devices operating at different wavelengths by acting on the input excitation; such applications would require a proper engineering of the cavity and will be the subject of future works.\\
%\paragraph{Conclusions. ---}
In conclusion, we  have developed   a  rigorous  theory   of  electromagnetic  energy-matter
interaction in the presence of  nonlinear resonant media. The derived equations are
applied  to   investigate  the  process  of  laser   emission  of  Mie
nanospheres.   By  means   of  parallel   simulations   and  frequency
demultiplexing  analysis,  we  discuss  both  the  transient  and  the
steady state regime of laser emission. During the transient, a
nontrivial mutual energy exchange  between the  electromagnetic wave
and  the  resonant medium occurs,  thereby  justifying a  rigorous  \emph{ab-initio}
investigation. On the other hand, the  lasing state  observed by increasing  the pumping
rate  is characterized  by a  dynamical transition  from  two-mode to
multi-mode  regime,  the   latter  accompanied   by  competition
phenomena and dynamical    frequency    pulling. Such results have important implications in the theory of
nano-lasers, in both ordered and  disordered systems  and are  expected to
stimulate  new  experiments and applications where  single Mie resonators  are  employed, 
including nonlinear optics, colloidal physics, chemistry and quantum mechanics.\\
%\paragraph{Acknowledgements. ---}
We acknowledge support from the INFM-CINECA initiative for parallel computing
and S. Trillo for useful discussions.
%\bibliography{../../refbib.bib}
%\bibliography{refbib.bib}

\end{document}